\documentclass[conference]{IEEEtran}
\IEEEoverridecommandlockouts

\usepackage{dirtytalk}
\usepackage{array}
\usepackage{tabularx}
\usepackage{graphicx}
\usepackage{subfig}
\usepackage{flushend}
\usepackage{hyperref}
\usepackage{amsmath,amssymb,amsfonts}
\usepackage{algorithmic}
\usepackage{graphicx}
\usepackage{textcomp}
\usepackage{xcolor}
\usepackage{subcaption}
\usepackage{cleveref}
\usepackage{ntheorem}
\theoremstyle{break}

\usepackage{algorithm}

\usepackage[backend=biber, style=ieee]{biblatex}
\begin{document}
\title{Maximal Extractable Value Mitigation Approaches in Ethereum and Layer-2 Chains: A Comprehensive Survey \\
\thanks{This work will be submitted to the IEEE for possible publication. Copyright may be transferred without notice, after which this version may no longer be accessible.}
}

\author{\IEEEauthorblockN{Zeinab Alipanahloo}
\IEEEauthorblockA{\textit{École de technologie supérieure} \\
Montreal, Canada \\
zeinab.alipanahloo.1@ens.etsmtl.ca}
\and
\IEEEauthorblockN{Abdelhakim Senhaji Hafid}
\IEEEauthorblockA{\textit{University of Montreal} \\
Montreal, Canada \\
ahafid@iro.umontreal.ca}
\and
\IEEEauthorblockN{Kaiwen Zhang}
\IEEEauthorblockA{\textit{École de technologie supérieure} \\
Montreal, Canada \\
kaiwen.zhang@etsmtl.ca}
}

\maketitle

\begin{abstract}
Maximal Extractable Value (MEV) represents a pivotal challenge within the Ethereum ecosystem; it impacts the fairness, security, and efficiency of both Layer 1 (L1) and Layer 2 (L2) networks. MEV arises when miners or validators manipulate transaction ordering to extract additional value, often at the expense of other network participants. This not only affects user experience by introducing unpredictability and potential financial losses but also threatens the underlying principles of decentralization and trust. 
Given the growing complexity of blockchain applications, particularly with the increase of Decentralized Finance (DeFi) protocols, addressing MEV is crucial. 
This paper presents a comprehensive survey of MEV mitigation techniques as applied to both Ethereum’s L1 and various L2 solutions. We provide a novel categorization of mitigation strategies; we also describe the challenges, ranging from transaction sequencing and cryptographic methods to reconfiguring decentralized applications (DApps) to reduce front-running opportunities. We investigate their effectiveness, implementation challenges, and impact on network performance. 
By synthesizing current research, real-world applications, and emerging trends, this paper aims to provide a detailed roadmap for researchers, developers, and policymakers to understand and combat MEV in an evolving blockchain landscape.
\end{abstract}


\section{Introduction} \label{INT}
The landscape of blockchain technology, particularly Ethereum and its Layer 2 (L2) chains, has been significantly impacted by MEV (Miner Extractable Value) strategies. These strategies, which include DEX (Decentralized Exchange) arbitrage, sandwiching, and liquidations, have emerged as profitable opportunities for miners and other participants in the network. However, these strategies also pose potential risks to the network's stability and security, leading to the development of various mitigation approaches. 
Blockchains have historically achieved data integrity and immutability through a network of nodes reaching consensus on the validity of transactions recorded in the distributed ledger. However, consensus protocols adopted in many blockchains do not enforce rules on the ordering of transactions produced by block producers. Miners or validators can exploit their capability to preview the network's upcoming state and manipulate transactions by reordering, including, or excluding them to generate profits. This profit-seeking behavior is known as MEV extraction; it can lead to significant issues for users, applications, and overall network robustness. This becomes significant with the rapid expansion of DeFi (Decentralized Finance) projects, leading to a notable increase in MEV extraction within the Ethereum network \cite{flashboys2.0}. 

One prominent approach to extract MEV is through front-running attacks; this involves submitting a transaction with higher transaction fees to surpass a pending transaction in the execution queue \cite{Shayan}. Front-running attacks form a dominant method to extract MEV; they enable individuals to leverage the transaction prioritization mechanism for their own financial gain.
It is important to note that transaction ordering manipulation also occurs on off-chain or layer 2 (L2) networks, where a single or multiple sequencer nodes are responsible for determining the final transaction order. L2 solutions are designed to improve the scalability and performance of blockchain networks by processing transactions off-chain \cite{paradigm_optimistic, mastersthesisTobias, surveyL2}.

Rollups are a specific type of L2 solutions with a single sequencer.
The sequencer, under the control of the rollup operator, is a central authority responsible for the ordering and execution of transactions.
Rollup chains employ a single sequencer to ensure deterministic transaction ordering, simplify system architecture, and enhance efficiency and security, although it may have certain drawbacks.
Rollups can also have a network of sequencers, though operating their own network can be costly for the Rollup chain.
\cite{rollupsHakim}.
A recent solution to this problem is the development of shared sequencer networks, which allow multiple Rollups to utilize a common network of sequencers.
This paper aims to survey the current state of MEV mitigation approaches on Ethereum and L2 chains, with a focus on the strategies employed to address the challenges posed by MEV. By grouping these mitigation approaches into specific strategies and analyzing them, we aim to provide a comprehensive understanding of the landscape of MEV mitigation. To the best of our knowledge, this work is novel in its systematic categorization and detailed analysis of MEV mitigation strategies.

The rest of this paper is organized as follows. Section \ref{RW} provides a brief overview of existing Surveys and reviews on MEV mitigation methods. Section \ref{BG} presents background concepts related to MEV extraction issue. Section \ref{TAX} introduces a novel taxonomy of different MEV mitigation strategies. Sections \ref{FO}, \ref{PP}, \ref{SCLP}, and \ref{PBS} review existing contributions for each mitigation strategy. Section \ref{OVER} compares all MEV mitigation strategies. Finally, Section \ref{CON} concludes the paper. 

\section{Related Work}\label{RW}
Previous research on MEV can be divided into two main categories. The first category involves the detection and quantification of MEV by analyzing recorded transactions or monitoring mempool activities, a topic extensively covered in the existing literature. The second category is concerned with strategies to prevent or mitigate MEV extraction within a network (e.g. \cite{buyingtime, Helix, blindperm, FairMM}). In this paper, we focus on the second category, specifically on MEV extraction mitigation and prevention methods on Ethereum and L2 chains.

Several studies have examined and evaluated various methods for mitigating MEV.
Xu et al. \cite{xu2103sok} focus on Automated Market Maker (AMM)-based Decentralized Exchange (DEX) platforms. They assess the mechanics of various protocols (e.g., Uniswap \cite{uniswap}, Curve \cite{curve}, Balancer \cite{balancer}, and DODO \cite{dodoex}). They conclude with a comprehensive review of the AMM-based DEX platforms, focusing specifically on their design and associated vulnerabilities including front-running.
The authors propose a taxonomy covering the economic risks and security issues associated with AMMs. 
The paper classifies front-running as a significant economic risk for AMM protocols within its risk taxonomy;  it evaluates the impact of these attacks on the economic dynamics of DEXs. It also proposes potential mitigation methods for front-running, such as enforcing sequencing rules and concealing transaction details. 

Alam et al. \cite{alam2024front} explore the central role of DeFi in the financial ecosystem, particularly focusing on security issues. The authors concentrates on front-running attacks, with less emphasis on mitigation strategies. 

Zhang et al. \cite{zhang2023combatting} focus on evaluating methods for detecting front-running vulnerabilities in smart contracts. The authors propose an algorithm to detect real-world front-running attacks in the Ethereum transaction history; the algorithm outperforms baseline methods. They introduce an approach to automatically collect vulnerable smart contracts that have been exploited in historical attacks. In addition, they put forward a benchmark of real-world front-running attacks, including the associated vulnerable smart contract code. The paper conducts an empirical evaluation of seven state-of-the-art front-running vulnerability detection tools using the constructed benchmark.
Eskandari et al. \cite{Shayan} propose a classification framework to understand different types of front-running attacks. They categorize these attacks into three main types: displacement, insertion, and suppression. The authors analyze various mitigation techniques (transaction sequencing, confidentiality, and smart contract design practices), evaluating their effectiveness and limitations. Additionally, the paper presents real-world examples of front-running attacks and their impact on blockchain systems and applications.
The authors in \cite{heimbach2022sok} classify and examine various schemes for mitigating transaction reordering manipulations. However, they do not cover recent advances and their categorization is not exhaustive.

After our review of existing surveys \cite{xu2103sok, alam2024front, zhang2023combatting, Shayan, heimbach2022sok}, we conclude that while they have looked into various strategies to mitigate MEV, few have taken the comprehensive and high-level approach that we aim to present in this paper. Most previous surveys focus on specific aspects of MEV mitigation, but they don't provide the complete picture. We aim to offer a more holistic perspective, covering a broader range of strategies and their interconnections.


\section{Background}\label{BG}
In this section, we introduce background concepts that are necessary to understand the rest of the paper. We begin by discussing MEV, explaining its significance and impact on blockchain transactions. Next, we present the details of front-running attacks, describing how they exploit transaction ordering to benefit malicious actors. Finally, we present L2 solutions highlighting how they enhance scalability in blockchain systems.

\subsection{MEV}

MEV \cite{flashboys2.0} refers to the maximum profit that a miner (miner and validator will be used interchangeably throughout the rest of the paper) can achieve from pending or confirmed transactions within a blockchain network. MEV represents the potential earnings validators can gain by strategically ordering transactions to their advantage. Validators possess the authority to include, exclude, or reorder transactions within a block. By leveraging this power, they can prioritize certain transactions that offer higher fees or manipulate transaction ordering to benefit from price movements, thereby maximizing their revenue. Currently, most actual MEV is extracted not only by validators but also by sophisticated users and automated bots, collectively known as MEV searchers. These searchers constantly monitor the blockchain for profitable opportunities, such as arbitrage, liquidations, and front-running. They employ advanced algorithms and high-frequency trading strategies to identify and capitalize on these opportunities before others can.
This practice often occurs at the expense of other users, leading to concerns about fairness and market efficiency within the blockchain ecosystem.
The implications for the users include: (1) financial loss: it happens through front-running on DEX platforms; (2) censorship: transactions can be censored or delayed by validators; and (3) high transaction fees: it is caused by network congestion.

\noindent MEV extraction also impacts the network layer, potentially destabilizing it through various attacks including (1) time-bandit attacks \cite{TimeBanditAttack}: validators attempt to rewrite blockchain history by reorganizing past blocks with profitable transactions; and (2) Priority Gas Auctions (PGA) \cite{flashboys2.0}: high transaction fees are bid to prioritize transactions and capture opportunities.

\subsection{Front-Running}

The most common method of extracting MEV is through front-running attacks. In these attacks, a transaction is placed ahead of another in the execution queue by bidding higher transaction fees. In Ethereum, transactions are temporarily stored in a public pool known as the mempool, which allows anyone to monitor and identify potential profit opportunities before they are included in a block. Malicious actors exploit this transparency by executing transactions with higher gas fees, effectively front-running the original transaction to capture profits before it is processed.
When multiple front-runners compete for the same opportunity, it leads to a PGA \cite{flashboys2.0}, where the transaction with the highest gas fee wins the race to be included in the next block. This competitive bidding process can escalate quickly, causing network congestion as more transactions compete for limited block space. As a result, the overall transaction fees on the network increase, impacting regular users who are not participating in the auction. 
In certain situations, a validator might choose to censor or delay a specific transaction. This could be to prioritize their own transaction in order to profit, particularly if the potential profit exceeds the combined block reward and transaction fees.

\begin{figure*}
  
\begin{minipage}[b]{0.5\textwidth}
   \subfloat[Rational Behavior of Miners or Validators]{\includegraphics[width=\columnwidth]{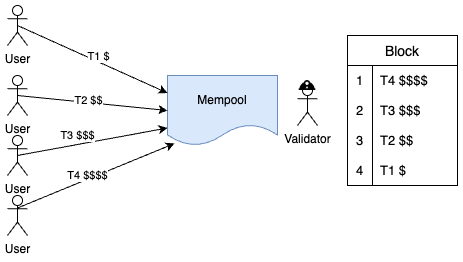}}\label{fig:rational}
\end{minipage}
    \hfill
 \begin{minipage}[b]{0.5\textwidth}
    \subfloat[Example of a Front-running Attack]{\includegraphics[width=\columnwidth]{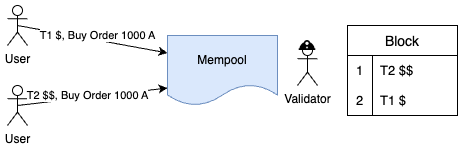}}\label{fig:frontrun}
\end{minipage}
    \hfill
 \begin{minipage}[b]{0.8\textwidth}
    \subfloat[Example of a Priority Gas Auction Attack]{\includegraphics[width=\columnwidth]{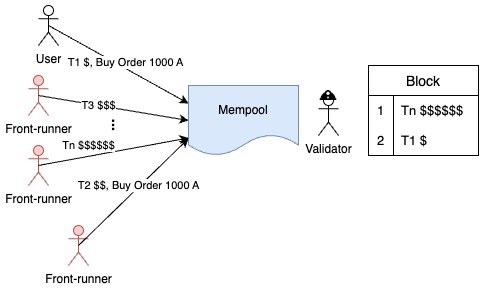}}\label{fig:PGA}
\end{minipage}
\caption[Delay Times]{
\label{fig:MEVEXAMpl}
Opportunity for Front-running Attacks Arising from validators' Arbitrary Transaction Selection and Ordering}
\end{figure*}

Fig. \ref{fig:MEVEXAMpl} (a) shows an example of a front-running attack. Users submit transactions with specific fees to a blockchain node, where they are stored in a pool of pending transactions known as a mempool. The validator selects transactions from the mempool to create a block; typically, miners prioritize processing transactions that offer higher fees.

Fig. \ref{fig:MEVEXAMpl} (b), the user plans to buy a significant amount of a certain token on a DEX (e.g. $1000$ token $A$). This large purchase will cause the token's price to rise after the order is executed. Before the transaction is broadcast to the network, a front-runner bot detects this large buy order in the mempool. The bot then places a buy order for the same amount of the token but with a higher transaction fee, allowing it to buy at a lower price. When the user's order is executed, it drives up the token's price, enabling the front-runner to sell their tokens at the increased price, thereby profiting from the user's transaction.

\paragraph{\textbf{PGA Example}}In the previous example, there was one front-runner who only needed to set a higher transaction fee than the user's specified fee.
Fig. \ref{fig:MEVEXAMpl} (c) shows a scenario where multiple front-runner bots detect the user's transaction. If the use's fee is $\$$, the first bot sets a higher fee, say $\$\$$. The next bot sees this and sets an even higher fee, $\$\$\$$. 
This competition among bots drives the transaction fees higher, creating a situation known as a \emph{Priority Gas Auction}. Meanwhile, the validator, aiming to maximize profit, waits to select the highest fee offers for inclusion in the block.

\subsection{L2 Scaling Solutions}

L2 solutions offer a way to scale the underlying chain or layer 1 (L1) chain. Each L2 protocol has its own execution environment, resembling the Ethereum
virtual machine. The key concept involves deploying decentralized applications (DApps) on L2 protocols, enabling computations and storage to take place off-chain. 
Rollups are a type of the L2 solutions, in which computations and the application's state are kept in an off-chain protocol, and the proof data of transactions are stored on L1. In this study, we focus on the rollup protocols as an efficient and general-purpose off-chain solution. 
Rollups aggregate multiple transactions into a single batch and perform most computation off-chain, using the underlying network for security.
There are two main types of Rollups. Optimistic rollups operate on the principle of assuming transactions are correct unless proven otherwise. In contrast, Zero-Knowledge rollups use cryptographic proofs to validate transactions.
A crucial component in the architecture of rollups is the sequencer. The sequencer plays a vital role in maintaining the correct order of transactions and ensuring the consistency of the network state. When users submit transactions to a rollup chain, these transactions are initially processed and validated off-chain within the rollup network. The sequencer then collects these transactions and orders them based on their received time or other criteria, creating a batch or block of transactions.
Periodically, the sequencer commits the batch to the underlying network, thereby ensuring the integrity of the transactions and the rollup’s state. Transaction ordering and execution can be separated in rollup protocols, but in practice, they can be gathered in a single node \cite{sokPatric, nitroArbitrum}.

\section{Taxonomy of MEV Mitigation Strategies}\label{TAX}
\begin{figure*}
\centerline{\includegraphics[scale=0.9]{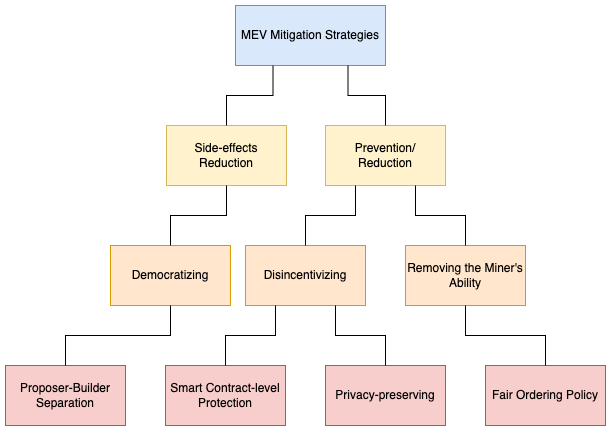}}
\caption{Taxonomy of MEV Mitigation Strategies}
\label{fig:taxonomy}
\end{figure*}

In this section, we present a novel taxonomy outlining various methods to mitigate MEV (see Fig. \ref{fig:taxonomy}). This taxonomy aims to offer a broad understanding of each mitigation technique, highlights its respective objectives and underlying strategies. 
The MEV mitigation taxonomy consists of two primary strategies, each addressing the issue from distinct perspectives: (1) prevention/reduction: it aims to eliminate MEV opportunities before they can be exploited; and (2) side-effect reduction: it focuses on mitigating the adverse impacts of MEV extraction rather than eliminating the opportunities themselves.

The prevention/ reduction strategy involves removing or limiting the validator's ability to reorder transactions, thereby directly preventing them from extracting MEV. Conversely, MEV searchers are unable to leverage higher transaction fees to exploit the sequencers' capabilities. Prevention can also be achieved through \textbf{Disincentivizing}; This involves creating deterrents against engaging in MEV practices. 

\noindent The side-effect reduction strategy focuses on mitigating the negative impacts of MEV extraction (PGA and time-bandit attacks) which can lead to network congestion and threaten the stability of the consensus protocol.
Instead of eliminating MEV opportunities entirely, this strategy focuses on democratizing MEV. By making MEV extraction accessible to a broader group of participants, the current imbalance where validators have a disproportionate advantage is reduced. 




In the following sections, we describe various proposed methods from the literature and industry, based on the strategies shown in Fig. \ref{fig:taxonomy}. 

\section{Fair Ordering Policy}\label{FO} 
A fair ordering policy is a method designed to eliminate a sequencer's ability to reorder transactions. It enforces a specific order for transactions to be included in a block by the sequencer. These policies are embedded into the protocol during its design by establishing explicit rules that the sequencer must adhere to when ordering transactions. These rules are enforced through the protocol's code and smart contracts, ensuring compliance.
While this approach can significantly reduce the likelihood of front-running, the sequencer node can still censor or delay transactions. Additionally, in the case of Rollups, where sequencer nodes have a private mempool, there is a risk that they could leak pool information to external front-runners, especially when there is only a single sequencer, which necessitates a strong trust assumption.

Ordering policies may be implemented in various settings. In a single sequencer scenario, a single node undertakes the sequencing task, with users directing their transactions to this node. Conversely, a multi-sequencer setup or decentralized sequencing involves multiple nodes collaborating to establish consensus on transaction order. The subsequent sections will delve into the challenges associated with fair ordering across these different configurations.
\subsection{Single Sequencer}
The most straightforward ordering approach applicable to a single sequencer node is the First-Come-First-Serve (FCFS) algorithm. It orders transactions according to their arrival time at the sequencer node. It is reasonable to prioritize transactions based on their arrival time, ensuring fairness in processing. Furthermore, this approach also reduces front-running by limiting sequencers' ability to order transactions arbitrarily. However, implementing FCFS in practice presents some challenges and drawbacks. These include (a) Latency wars: users try to optimize their network latency with sequencer nodes to guarantee early inclusion of their transactions to a block; and (2) Spam attacks: malicious users flood the sequencer node with a large number of transactions with the intent to guarantee their priority in the queue. As a result, legitimate transactions may experience delays or even fail to be processed altogether.


\begin{table}
\caption{Prominent Rollup protocols with their sequencing model}
\label{tab:rollups}
\begin{tabularx}{0.50\textwidth} { 
  | >{\centering\arraybackslash}X 
  | >{\centering\arraybackslash}X 
  | >{\centering\arraybackslash}X 
  | >{\centering\arraybackslash}X | }
 \hline
 \textbf{Protocol} & \textbf{Rollup Type} & \textbf{Sequencing} & \textbf{Ordering Policy} \\
 \hline
 \hline
 Arbitrum  & Optimistic & Single  & FCFS \\
\hline
 Optimism  & Zero-knowledge  & Single & FCFS \\
\hline
 Starknet  & Zero-knowledge & Single  & FCFS  \\
\hline
 Zksync  & Zero-knowledge & Single  & FCFS  \\
\hline
 Scroll  & Zero-knowledge & Single  & FCFS  \\ 
\hline
 Metis  & Hybrid & Decentralized  & FCFS  \\ 
\hline
\end{tabularx}
\end{table}

Most notable Rollups utilize a single sequencer with a FCFS ordering policy (see Table \ref{tab:rollups}). FCFS ordering policy simplifies management and implementation processes but has certain implications. Firstly, users must trust the sequencer node operated by the company. Secondly, there is a risk of a single point of failure; if the sequencer node encounters issues or is compromised, it could disrupt the entire Rollup system, causing transaction delays or failures.
\begin{figure}
\centerline{\includegraphics[width=\columnwidth]{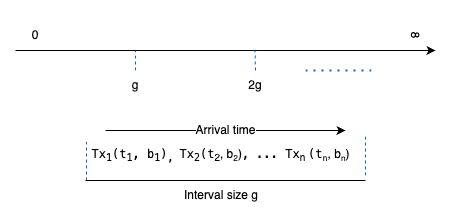}}
\caption{TimeBoost Algorithm}
\label{fig:granularity}
\end{figure}
Some protocols employ modified versions of the traditional FCFS algorithm to improve the fairness of transaction ordering. Although these solutions still contend with trust and single point of failure concerns, they can address issues associated with classic FCFS ordering, such as transaction spamming and latency optimization problems.


Kelkar et al. \cite{buyingtime} introduce \textbf{TimeBoost}, a fair ordering protocol in a single sequencing setting. Fig. \ref{fig:granularity} shows the proposed algorithm, which orders transactions within a granularity interval \( g \) on Rollup sequencers. It combines transaction timestamps and user bids to generate a score that determines the transaction order. Users can bid to effectively reduce their transaction's timestamp in the queue, effectively buying time. During the interval \( g \), transactions are initially sorted based on their received time \( t_i \); however, each user can bid \( b_i \) with their transaction to decrease the received time, thereby increasing the transaction's score. Ultimately, transactions are ordered based on their computed scores.
The score for a transaction \( tx_i = (t_i, b_i) \) is computed as follows:

\[ S(t_i, b_i) = \pi(b_i) - t_i \]

The function \( \pi(b_i) \) represents the priority achieved by bidding \( b_i \). Transactions are ordered by descending scores; a higher score increases the likelihood of the transaction being processed sooner. The bidding function \( \pi \) is designed to satisfy the following properties:
\begin{enumerate}
    \item \( \pi(0) = 0 \): Paying a bid of 0 provides no additional advantage.
    \item \( \pi'(b) > 0 \): For all \( b \in \mathbb{R}^+ \), the priority increases with the bid.
    \item \( \lim_{b \to \infty} \pi(b) = g \): No transaction can outbid a transaction that arrived \( g \) time units earlier.
    \item \( \pi''(b) < 0 \) for all \( b \in \mathbb{R}^+ \): The priority function is concave, meaning the cost of gaining priority increases with the bid.  
\end{enumerate}

The simplest bidding function that satisfies these constraints is expressed as follows:

\[ \pi(b_i) = \frac{g b_i}{b_i + c} \]
%
Where $c$ is constant. Unlike FCFS, TimeBoost avoids the inefficient latency competition inherent to FCFS policies. By incorporating bids into the transaction ordering process, it encourages players to focus on bidding rather than investing heavily in low-latency infrastructure. In pure-latency approaches like FCFS, only the fastest players benefit, often leading to unfair advantages.

\noindent TimeBoost combines timestamps and bids, allowing bidding between transactions within a short time frame. This makes it superior to pure-bidding approaches, which can lead to attacks (e.g., PGAs), that can endanger network stability.
%

\subsection{Decentralized Sequencing}
Decentralized sequencing has emerged as a viable alternative by addressing the limitations associated with relying on a single sequencer (e.g., trust and single point of failure). However, it introduces a substantial challenge: accurately determining the transaction order. In decentralized systems, this frequently involves the need for synchronized clocks. Because of network delays, influenced by the geographical distances between nodes and the available network bandwidth for users, transactions might reach different nodes at different times and in various orders. As a result, the transaction order in each node's mempool may vary. 


\textbf{Metis} \cite{Metis2024} is one of the recent Rollup chains pioneering decentralized sequencing, primarily addressing the single point of failure issue. It features a straightforward method for ordering transactions in a decentralized environment. 
All sequencer nodes receive transactions, and in each round, one sequencer is selected to construct a block.
Metis uses a rotation mechanism for Sequencer selection, with all Sequencer Lists stored in a smart contract address controlled by multiparty computation (MPC). The selection is based on each Sequencer node's voting weight (related to the staked amount of token) and a randomly generated hash value, ensuring a fairer process. If a Sequencer stops operating, it is rotated out to maintain continuous network operation. 
The selected sequencer orders transactions based on the FCFS policy using the order of transactions in its local mempool.
Metis has a permissioned pool of sequencers governed by a DAO (Decentralized Autonomous Organization). Prospective sequencers must submit proposals, subject to a voting process requiring majority approval from DAO members to join the network. Additionally, sequencers are mandated to stake a certain quantity of tokens as a deterrent against malicious behavior. Mechanisms for slashing exist to penalize sequencers, while rewards are offered for transaction processing and block production, thereby incentivizing honest behavior.
The slashing mechanism addresses malicious actions or poor performance by Sequencer nodes. Offending Sequencers are immediately removed from the pool and subjected to a review process. Slashing cases, categorized by severity, include slow Sequencers (low severity), non-block production (medium severity), multiple node outages (high severity), and malicious execution result modification (critical severity). Programmatic detection and enforcement handle these issues, with additional penalties determined by governance proposals. Actions against offenders can include slashing a percentage of their stake, blacklisting, and removal from consensus \cite{Metis2024, metisGOV2024}.


More complex forms of decentralized sequencing have been proposed in the literature including \cite{themis, wendy, kelkar2020order, zhang2020byzantine}; however, they have not yet been implemented in a production blockchain. In these methods, all sequencer nodes receive transactions, and a leader sequencer is selected each round to establish the final transaction order. Instead of relying solely on its local mempool, the leader integrates the local orders from all sequencers based on a fair ordering definition. This definition determines the final order of transactions by deciding, for each transaction pair \((Tx_0, Tx_1)\), which one should be processed first.

\begin{figure*}
    \centering
    \includegraphics[width=1\textwidth]{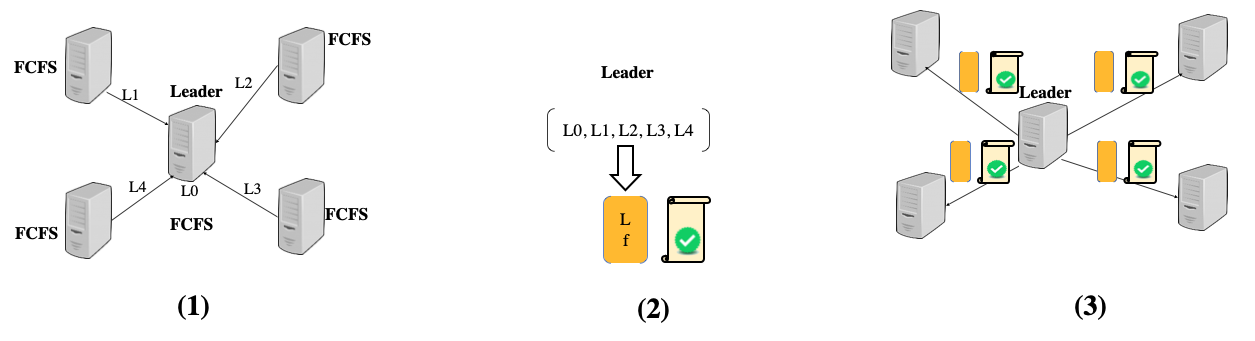}
    \caption{Themis Network Protocol}
    \label{fig:themisProtocol}
\end{figure*}

Kelkar et al. \cite{themis} introduced \textbf{Themis} which provides a fair ordering of transactions. Fig. \ref{fig:themisProtocol} shows the Themis protocol ordering process. It is a leader-based protocol where all nodes transmit their transaction orders to the chosen leader, determined by their receive times (see Fig. \ref{fig:themisProtocol} (1)). Subsequently, the leader compiles a fair-ordering proposal from the transaction orders received from other nodes (see Fig. \ref{fig:themisProtocol} (2)). Finally, the leader sends the final ordering, accompanied by a SNARK proof of computation, to all other sequencer nodes involved in the sequencing process (see Fig. \ref{fig:themisProtocol} (3)).
The fair ordering definition used by Themis is known as \emph{Batch-Order-Fairness}. According to this definition, if two transactions $tx_0$ and $tx_1$ are received by all sequencer nodes, and if a fraction (more than $\frac{1}{2}$) of these nodes received $tx_0$ before $tx_1$, then all honest nodes should order $tx_0$ no later than $tx_1$. 
A dependency graph is created where the vertices represent transactions and directed edges indicate the order between pairs of transactions. Additionally, a proof mechanism allows nodes to verify the accuracy of the final ordering proposed by the leader, based on the specified fair ordering algorithm \cite{themis}. This ordering policy can encounter the phenomenon known as the \textbf{Condorcet paradox or cycle}. 

\begin{figure}
\centerline{\includegraphics[width=\columnwidth]{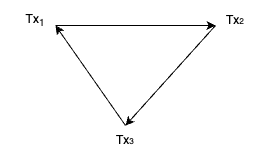}}
\caption{Example of a Condorcet Cycle}
\label{fig:condorcet}
\end{figure}

\noindent\textbf{For example}, let us consider three transactions received by three sequencer nodes, with each node receiving the transactions in a different order. The orders are presented as follows: 
\begin{itemize}
    \item \textbf{Node A}: \(Tx_1>Tx_2>Tx_3\)
     \item \textbf{Node B}: \(Tx_2>Tx_3>Tx_1\)
      \item \textbf{Node C}: \(Tx_3>Tx_1>Tx_2\)
\end{itemize}
Fig. \ref{fig:condorcet} shows the resulting dependency graph.
When cycles are present in the graph, it becomes impossible to determine the final order of transactions within the cycle. Moreover, the consensus process can be stalled. In protocols like Aequitas \cite{Aequitas}, this leads to a loss of liveness, where transactions are not processed in a timely manner or at all; the protocol gets stuck trying to resolve the cycle.
High latency in a network can cause nodes to receive transactions in different orders, increasing the likelihood of conflicting preferences and thus cycles.
Themis addresses this issue by using a method called \emph{deferred ordering}, which helps manage the impact of cycles by ensuring that transactions do not have to wait indefinitely for ordering. 
In the following, we present the steps executed by Themis in the context of the example shown in Fig. \ref{fig:condorcet}. Let us consider that node $A$ is selected as a leader.

\begin{enumerate}
    \item \textbf{Partial Ordering: }Node $A$ proposes a block that includes transactions $Tx_1$ and $Tx_2$ as fully ordered, but $Tx_3$ is only partially ordered. This means that the final position of $Tx_3$ in the block is not yet determined.
    \item \textbf{Deferred Ordering Process: } The proposed block is broadcast to the other nodes. Nodes $B$ and $C$ receive the block from Node $A$. They agree on the order of $Tx_1$ and $Tx_2$ but still have their own views on $Tx_3$.
    The nodes defer the final ordering of $Tx_3$ to the next block proposed by the next leader.
    \item \textbf{Final Ordering by the Next Leader: } Node $B$, the next leader, creates a new block. This block finalizes the position of $Tx_3$, considering the partially ordered state from node $A$'s block and any new transactions.
    Node $B$ includes a complete ordering for $Tx_3$ along with new transactions (e.g., $Tx_4$, $Tx_5$).
    \item \textbf{Consensus and Block Commitment: }The new block proposed by node $B$ is broadcast and agreed upon by all nodes. The blockchain now has a consistent order: Block 1 ($Tx_1$, $Tx_2$, partially ordered $Tx_3$), Block 2 ($Tx_3$, $Tx_4$, $Tx_5$).
\end{enumerate}

Kursawe \cite{wendyGrowsUp} introduces Wendy which applies timestamp-based ordering in a decentralized setting; they leverage synchronized clocks to ensure that transactions are processed in the order they are received based on their timestamps. In Wendy, all nodes in the network maintain synchronized clocks. This synchronization can be achieved using protocols such as Network Time Protocol (NTP). When a node receives a transaction, it assigns a timestamp based on its local clock. The node then broadcasts the transaction along with its timestamp to the rest of the network. Each node has a mempool where it stores incoming transactions along with their timestamps. Nodes follow a predefined ordering policy based on the timestamps. Transactions with earlier timestamps are given higher priority for inclusion in the next block. During the consensus process, nodes propose and validate blocks of transactions. The validator must order the transactions in the block according to their timestamps, ensuring that transactions with earlier timestamps are included before those with later timestamps. When a node receives a proposed block, it verifies that the transactions are ordered correctly according to their timestamps. If the ordering is not correct, the block is rejected.   

\subsection{Shared sequencers}
Shared sequencing is a technique that has been discussed in recent years within the Rollup ecosystem. It offers transaction sequencing for multiple Rollups through a third-party service, enabling the sequencing process to be shared among various Rollup chains. 
A shared sequencer operates independently of the Rollup chain by separating the executor role from the sequencer role. In this setup, the sequencer functions solely as an sequencer. A shared sequencer can either be a single sequencer or a decentralized one. In the latter case, similar to a network of validators in a blockchain, the shared sequencing network forms a peer-to-peer network among sequencer nodes. It can significantly reduce the cost of running Rollups and accelerate the move towards decentralization. However, Rollups will need to share transaction fees and MEVs with other entities.

\begin{figure}
\centerline{\includegraphics[width=\columnwidth]{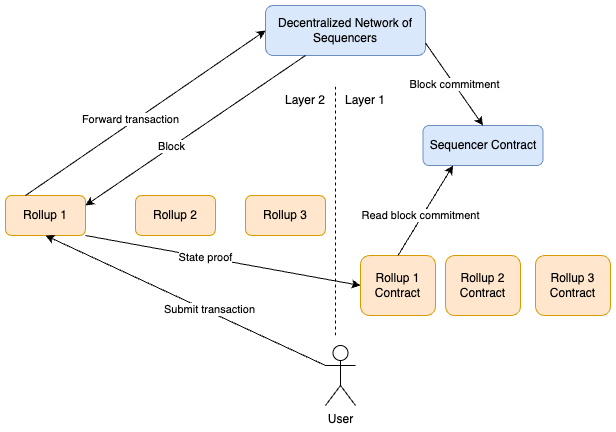}}
\caption{Espresso Network Architecture as a Decentralized Shared Sequencer}
\label{fig:espresso}
\end{figure}

\textbf{Espresso} \cite{espressoSequencer} is an example of a shared sequencing protocol. The architecture of the Espresso system is illustrated in Fig. \ref{fig:espresso}. When a user sends a transaction to a Rollup chain, following steps are executed:

\begin{enumerate}
    \item The Rollup receives the transaction and forwards it to the shared sequencer.

    \item The sequencer network creates blocks containing an ordered list of transactions.
    \item The Rollup receives the block and executes transactions on its network.
    \item The sequencer posts the log of executed transactions to L1 as a commitment (on its sequencer smart contract).
    \item After executing the block of transactions, the Rollup posts the updated state of the chain to L1 (on the Rollup smart contract).
    \item The Rollup contract then reads the block commitment from the sequencer contract. The verification of state transition depends on the Rollup type: a ZK-Rollup verifies the proof, while an optimistic Rollup waits for fraud proofs.
\end{enumerate}
This process occurs simultaneously for multiple Rollups. There is an incentive for Rollups to forward transactions to the Espresso network, known as the sequencer marketplace. When a Rollup receives a transaction from a user, it can choose to process it through its own sequencer node or sell its sequencing rights through an open marketplace.
\cite{EspressoSys2024}. 

 \subsection{Discussion}
 Table \ref{tab:deSequener} presents a comparison of notable protocols discussed in the previous sections on fair ordering methods. Most protocols feature a set of permissioned sequencers, except for Espresso, \cite{EspressoSys2024} which allows anyone to join the network and run their own sequencer node. All the protocols are leader-based, with a single sequencer acting as the leader or, in the context of decentralized sequencing, one sequencer being selected as the leader for each block.
All the protocols mentioned in the table, except Wendy \cite{wendyGrowsUp}, do not require a synchronized clock for their algorithms. Wendy’s use of timestamps for each transaction increases communication overhead. Over time, clocks on different nodes can drift apart due to slight differences in their frequencies, leading to inconsistencies in timestamps and potential issues in the fair ordering process. Synchronizing clocks also adds network traffic, as nodes must periodically exchange time information, consuming additional computational and bandwidth resources that could be used for other operations.

Table \ref{tab:rollups} indicates that most Rollups use the classic FCFS ordering policy. TimeBoost \cite{buyingtime} is a proposed transaction ordering method for single sequencer Rollups, combining bidding and time-based ordering methods.
Metis is another Rollup network that has proposed a decentralized sequencing method. While it addresses the single point of failure issue and reduces the trust required in a single sequencer, it still relies on the classic FCFS sequencing method.
Themis \cite{themis} and Wendy \cite{wendyGrowsUp} both propose decentralized sequencing methods but with different ordering policies. Themis operates without requiring a synchronized clock and maintains an $O(n)$ communication cost between sequencer nodes, utilizing the SNARK protocol. In contrast, Wendy requires synchronized clocks and has an $O(n^2)$ communication complexity.
Shared sequencing protocols like Espresso \cite{EspressoSys2024} reduce the cost of running a network of sequencers for Rollup chains. Rollups can either use their own sequencing policy or delegate the task to a third-party sequencing network through an auction mechanism.

\noindent As Wendy uses a timestamp-based ordering mechanism; making it difficult for any single node to censor transactions because the order is determined by the timestamps, not by any single node’s decision.
Themis enforces batch-order fairness, meaning that transactions are ordered based on the order in which they were seen by a significant fraction of honest nodes. This reduces the ability of any single node to manipulate the order for censorship purposes. It employs SNARKs to achieve efficient and secure validation of the ordering process. This ensures that even if a node tries to censor transactions, it cannot tamper with the verification process without being detected.


\begin{table*}
\caption{Notable Sequencing Protocols}
\label{tab:deSequener}
\begin{tabularx}{1\textwidth} { 
  | >{\centering\arraybackslash}X 
  | >{\centering\arraybackslash}X 
  | >{\centering\arraybackslash}X 
  | >{\centering\arraybackslash}X
  | >{\centering\arraybackslash}X
  | >{\centering\arraybackslash}X
  | >{\centering\arraybackslash}X
  | >{\centering\arraybackslash}X
  | >{\centering\arraybackslash}X | }
 \hline
 \textbf{Protocol} & \textbf{Sequencing Model} & \textbf{Permissioned} & \textbf{Leader based} & \textbf{Synchronized Clock} & \textbf{Ordering Policy} & \textbf{Verification} & \textbf{Censorship Resistance} & \textbf{Comm. Complexity}\\
 \hline
 \hline
TimeBoost \cite{buyingtime}  &Single & Yes& Yes  & No & Modified FCFS  & Trust & No & ---\\
\hline
 Metis \cite{Metis2024}  & Decentralized & Yes& Yes  & No & FCFS  & DAO, Slash & No & ---\\
\hline
 Themis \cite{themis} & Decentralized & Yes & Yes& No& Batch-order fairness  & SNARK & Yes & $O(n)$ \\
\hline
 Wendy \cite{wendy} & Decentralized & Yes & Yes & Yes & Timestamp-base  & Timestamps & Yes & $O(n^2)$\\
\hline
 Espresso \cite{EspressoSys2024} &Shared & No & Yes & No & Auction-based  & PoS consensus, Slashing Rules & Yes & O(n)\\
\hline
\end{tabularx}
\end{table*}

\section{Privacy-preserving}\label{PP}
The main objective for the privacy-preserving methods is to hide the contents of transactions until their order is confirmed. These approaches can effectively eliminate MEV extraction opportunities and reduce the risk of transaction censorship by malicious sequencers. Since the transaction is not visible in the public mempool, it is less vulnerable to front-running and other types of attacks that exploit transaction visibility.

\begin{figure}
\centerline{\includegraphics[width=\columnwidth]{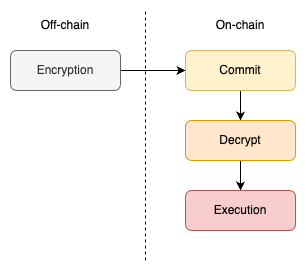}}
\caption{Mempool Encryption General Approach}
\label{fig:memEnc}
\end{figure}

\noindent The primary focus of research in this domain is on \textbf{mempool privacy} or \textbf{encrypted mempool} (see Fig. \ref{fig:memEnc}), involving the encryption of pending transactions in the mempool and their decryption after their order is finalized, just before inclusion in a block. 
\cite{encryptedmempools2023}.

Another privacy-preserving approach that helps prevent front-running involves \textbf{private transactions}. This is achieved by bypassing the mempool and directly sending your transaction to a validator, who then places the transaction in the newly created block ahead of other transaction \cite{empiricalprivate}. 

    
\subsection{Encrypted Mempool}

The main techniques for encrypting the mempool include (1) \textbf{Threshold encryption}, which entails multiple trusted parties (referred to as \emph{key holders} or \emph{keypers}) decrypting transactions collaboratively without revealing the decryption key; (2) \textbf{Delay encryption}, where transactions remain encrypted for a designated duration before decryption; and (3) \textbf{Trusted Execution Environments (TEEs)} which guarantee security and privacy prior to transaction confirmation \cite{unlockingmempool}. We will explore these three methods in detail.

\subsubsection{Threshold Encryption}
\begin{figure}
\centerline{\includegraphics[width=\columnwidth]{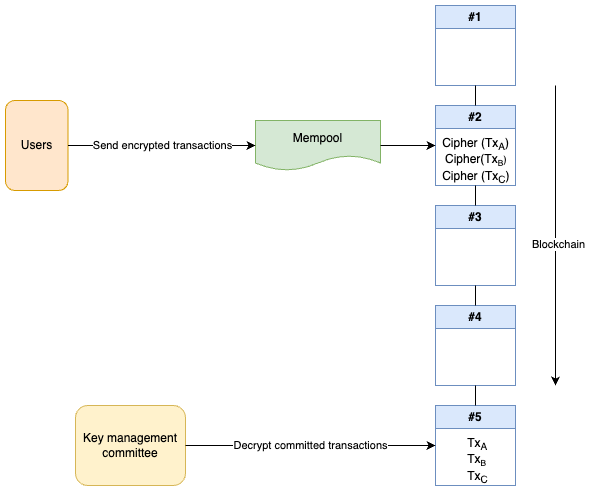}}
\caption{Sending Encrypted Transactions by the User Using the Threshold Encryption Method}
\label{fig:thresholdEnc}
\end{figure}

\hfill \break
In threshold encryption, the symmetric key is divided into several pieces (shares) and distributed to a key-holding committee during the encryption stage. To decrypt, a certain threshold of the total shares is required to reconstruct the symmetric key,  making it impossible for a single attacker to attempt decryption before the transaction order is decided. This requires a degree of trust in the key-holding committee as a third party. The trust assumption here is that a certain threshold of committee members will share their portions of the key in a timely manner. The general process of the threshold encryption method is outlined in the following steps (see Fig. \ref{fig:thresholdEnc}).

\begin{enumerate}
    \item \textbf{Key generation:} Each member of the committee generates a share of the decryption key based on the Distributed Key Generation (DKG) algorithm \cite{DKGAlgorithm}. This process ensures that no single member has the complete decryption capability, thereby distributing control over the decryption process among multiple parties.
    \item \textbf{Encryption phase:} Transactions are encrypted using a public key from the trusted key management committee with an honest majority
    \item \textbf{Submission of encrypted transactions:} Encrypted transactions are submitted to the mempool. Since the transactions are encrypted, they cannot be tampered with or manipulated by validators.
    \item \textbf{Transaction commitment:} The encrypted transactions are then posted on the blockchain as a commitment, but they can not be executed. At this point, the transactions are visible to all participants, but they are encrypted, and their contents are not accessible without the decryption key.
    \item \textbf{Decryption phase:} Once a transaction is recorded on the blockchain, the key holders can initiate the decryption process. The contents of the transactions remain private until a threshold number of holders start the decryption process. This safeguard ensures that, even in the event of some participants being compromised, the integrity of the transactions is maintained as long as the majority of participants remain honest.
    \item \textbf{Execution of Transactions: }Once the threshold decryption process concludes, the transactions are processed in a sequence based on the commitment already posted on the blockchain. This predetermined order is unrelated to the content of the transactions, effectively blocking any potential for MEV exploitation through strategies that rely on content-based ordering.
\end{enumerate}

The key management committee can be chosen from validators on the main chain, who are responsible for consensus and block creation \cite{ferveo}. 
However, this approach may introduce overhead during the decryption phase and necessitate modifications to the protocol layer. Alternatively, an off-chain mechanism managed by trusted parties can handle key generation and decryption processes. An example of such a protocol is \textbf{Shutter} \cite{shutter}, which offloads the key management process to an off-chain environment.
Another key advantage of Shutter, in terms of efficiency, is that it uses a single decryption key for each epoch in POS consensus (approximately for 32 blocks), whereas Ferveo generates a decryption key for each transaction. \cite{choudhuri2024mempool}.



Yakira et al. \cite{F3B} propose Flash Freezing Flash Boys (F3B), a novel blockchain architecture that employs a commit-and-reveal scheme. In this scheme, the contents of transactions are encrypted and only revealed by a decentralized secret-management committee after the underlying consensus layer has committed the transaction. 
The key components of F3B include: encrypted transaction senders, consensus group, and secret-management committee (set of trusted nodes which are responsible for keeping the encryption key). Although consensus nodes and management nodes are two separate entities in some situations (e.g., when a permissioned blockchain is used as the underlying blockchain) they can be the same or can be run on a single server.
The  process of the F3B threshold encryption method is outlined in the following steps:
\begin{enumerate}
    \item \textbf{Key Generation Phase:} The secret-management committee generates a secret key for each member and a single public key using a DKG algorithm \cite{DKGAlgorithm}. This process is repeated at specific intervals during the \emph{reconfiguration} phase, which allows for the addition and removal of committee members. 
    \item \textbf{Key Selection and Encryption Phase:} 
    \begin{itemize}
        \item \textbf{Symmetric key generation:} The user or the sender of the transaction generates a random number on their side; it is considered as the symmetric key $k$.
        \item \textbf{Encryption of Symmetric Key:} The sender then encrypts the symmetric key $k$ using the public key $pk_{smc}$  of the secret-management committee to produce the encrypted symmetric key $ck$.
        \item 
        \textbf{Transaction Encryption:} The sender encrypts the transaction $tx$ using the symmetric key $k$, resulting in the encrypted transaction $c{tx}$.
        \item \textbf{Sending Encrypted Data:} The sender sends the pair $(c_{tx}, c_k)$ to the consensus group, which validates the integrity of the data. Once validated, the pair is included in a new block of the blockchain.  
    \end{itemize} 
    
    \item \textbf{Waiting for Commitment:} At least $m$ block confirmations are needed for the commitment phase. 

    \item \textbf{Key Release and Execution:}
    \begin{itemize}
        \item \textbf{Retrieving the Encrypted Symmetric Key:} Each member in the secret-management committee retrieves the encrypted symmetric key $c_k$ from the blockchain. The key $c_k$ was previously encrypted with the committee's public key $pk_{smc}$ and recorded on the blockchain during the second phase. Each member individually decrypts their share of the symmetric key $k$. The decryption is done using the member's private key share, which corresponds to their share in the DKG algorithm used to generate $pk_{smc}$. This results in partial decrypted key shares.
        \item \textbf{Verification by Consensus Nodes:} To ensure that the decryption was done correctly, each member generates a non-interactive zero-knowledge (NIZK) proof. This proof verifies that their decrypted key share is correct without revealing the key share itself. The consensus nodes in the blockchain network verify the NIZK proofs provided by the trustees. If the proofs are valid, it confirms that the decrypted key shares are correct and trustworthy.
        \item \textbf{Reconstructing the Symmetric Key:} Once a sufficient number of correct key shares and corresponding proofs are collected, the consensus nodes use Lagrange interpolation to reconstruct the original symmetric key $k$. This is possible because the DKG algorithm ensures that the key can be reconstructed with a threshold number of shares.

        \item \textbf{Decrypting the Transaction:} With the symmetric key $k$ reconstructed, the consensus nodes decrypt the encrypted transaction  $c{tx}$ to obtain the original transaction $tx$.

        \item \textbf{Transaction Verification and Execution:} The decrypted transaction $tx$ is then verified to ensure it meets all necessary conditions and is valid. Once verified, the transaction is executed according to the rules and logic defined in the blockchain protocol.
        \end{itemize}
        
\end{enumerate}

F3B requires writing data onto the blockchain only once, significantly reducing overhead. 
However, F3B requires modifications to the blockchain protocol to enable consensus nodes to verify and process encrypted transactions and key shares; Additional logic must be added to handle the decryption phase and the subsequent transaction validation \cite{F3B}.

\textbf{Helix} \cite{Helix} is a hybrid protocol that combined the idea of fair ordering with threshold encryption. The protocol operates on a fully connected and synchronous network of nodes, ensuring that each transaction propagates to all nodes, resulting in all nodes having the same set of transactions. \textbf{Helix} is a leader-based protocol where the leader is selected among a committee of nodes, and the committee itself is established through an election process.
Each node has a reputation score that increases with honest participation in block validation and proposals. The higher a node's reputation score, the greater its chances of being selected as a committee member or leader.  

As the consensus and transaction ordering processes are confined to a limited number of committee nodes, the protocol scales with the number of nodes.
The protocol selects a random set of pending encrypted transactions to include in a block using a Verifiable Random Function (VRF) \cite{verifiable}; the objective is to guarantee that the selection of pending transactions is done in a transparent, verifiable, and unbiased manner, enhancing the fairness and security of the blockchain network.
It ensures transaction censorship resistance by employing a threshold encryption mechanism, preventing committee and leader nodes from knowing the content and the issuer address of a transaction until their order is finalized.

For example, if Node $P$ is selected to propose a block in one round among the committee members, the process of block creation is as follows.
\begin{enumerate}
    \item A random seed is generated for the round; it is used as the input for the VRF. 
    \item Node $P$ computes a VRF output using the random seed and its private key share. It also generates a proof of VRF computation that can be verified by other nodes without knowing the private key.
    \item Each transaction in the pool is hashed to generate a unique identifier.
    For each transaction $tx_i$, node $P$ concatenates the transaction’s hash with the VRF output to form a combined string. The combined string is then hashed to produce a ranking score for the transaction.
    \item The transactions are ordered based on their ranking scores in an ascending order. This ensures that each node, using the same VRF output and the same set of transactions, will compute the same order for the transactions.
    \item Node $P$ creates a block with the top $k$ transactions; along with VRF proof, the block is proposed to other committee members.
    \item The committee members verify the VRF proof to ensure that the transaction selection process was random and fair. They run a Byzantine Fault Tolerance (BFT) protocol to reach consensus on the proposed block.
    \item Once consensus is reached, nodes begin the threshold decryption process for the transactions in the block. The decrypted transactions are then validated for correctness.
    \item The validated transactions are included in the new block, which is then appended to the blockchain.

\end{enumerate}


mempool Encryption can be improved with other techniques. For example, Kavousi et al. \cite{blindperm} introduce a framework called \textbf{BlindPerm}, which enhances encrypted mempool, with permutation techniques, providing multi-layer protection against MEV. It uses randomized permutations to shuffle the order of transactions within a committed block before they are executed. This framework is designed to work with BFT-style consensus mechanisms and is shown to be efficient; it introduces essentially no overheads and requires no additional services. The process of the BlindPerm is outlined in the following steps.
\begin{enumerate}
    \item The validator who is selected as the leader, proposes a block , containing encrypted transactions with an arbitrary order from the mempool.
    \item Once a block is committed, validators decrypt the transactions, based on the used mempool encryption, and derive a seed for permutation.
    \item Each validator uses the derived seed to permute the order of transactions by applying the permutation function $Permute(seed_i, B_i)$ to shuffle the transactions in the block.
    \item The permuted block $B'_i$ is executed by all validators.
\end{enumerate}

\subsubsection{Delay Encryption}
\hfill \break
\textbf{Delay encryption}, also known as \textbf{time-lock encryption}, is a cryptographic mechanism which provides a secure way to ensure that data remains confidential until a specified time has passed. It leverages cryptographic primitives like Verifiable Delay Functions (VDFs) and Time-lock Puzzles. This concept is particularly useful in scenarios where data should remain confidential until a specific future date or a condition is met. The core idea behind delay encryption is to tie the decryption key to a time-dependent condition (time-lock puzzle), making it inaccessible until that condition is satisfied.

VDF is a cryptographic primitive that allows a prover to convince a verifier that they have waited for a certain amount of time without revealing any information about the time they waited. This property is crucial for delay encryption; it allows the decryption key to be derived from a time-dependent condition without revealing the actual time waited.
Time-lock puzzles are designed in such a way that solving them requires a certain amount of computational work or time. The solution to the puzzle, which is the decryption key, can only be obtained after the required time has passed \cite{delayencryption, buildTimeLockEncryption}. For better understanding, let us consider the following example.

\begin{itemize}
    \item \textbf{Creating a Time-lock Puzzle:}
    \begin{enumerate}
        \item Generate prime numbers: \(p=61, q=53\)
        \item Compute \(N= p\times q\): \(N= 61 \times 53= 3233\)
        \item Choose a generator number and an exponent: \(g=2, t=1000\)
        \item Compute puzzle \(g^{2^t} mod N\): \({2^{(2^{1000})}} mod {3233}\)
        \item Drive a symmetric key \(K\) by hashing the result of the computation which is the puzzle solution \(S_o\): \(K=H(S_o)\)
        \item Encrypt a transaction with \(K\): \(C_{T_x}=Enc(K, T_x)\)
        
    \end{enumerate}
    \item \textbf{Solving a Time-lock Puzzle:}
    \begin{enumerate}
        \item Solve the time-lock puzzle using public parameters \(N, g, t\) by raising \(g\) to the power of \(2^{t}\). This process takes a specific amount of time, thereby introducing a delay before the decryption phase.
        \item Drive the symmetric key \(K\) by hashing the computed solution from the previous step: \(K=H(S_o)\)
        \item Decrypt the transaction with key \(K\): \({T_x}=Dec(K, C_{T_x})\) 
    \end{enumerate}
\end{itemize}

The  process of Radius is outlined in the following steps.
\textbf{Radius} \cite{Radius2024} is a shared sequencing protocol designed to enhance interoperability among Rollup chains. It employs delay encryption to create an encrypted mempool within its sequencing layer. This approach helps prevent MEV extraction through front-running and other reordering attacks, while also ensuring censorship resistance. 
The process starts with generating a transaction by a user, the following steps are as follows.
\begin{enumerate}
    \item The user generates a symmetric key using a time-lock puzzle set to a specific time \( T \). While the user already knows the solution, it will take the sequencer layer time \( T \) to solve the puzzle and reconstruct the symmetric key.
    \item The user encrypts the transaction using the generated symmetric key.
    \item The user also generates a ZK-SNARK proof for the sequencing layer to verify the integrity of the time-lock puzzle and the encrypted transaction.
    \item The user sends the encrypted transaction and the proof to the sequencing layer.
    \item Upon receiving the encrypted transaction, the sequencing layer provides users with an order commitment, ensuring the transaction's sequence remains unchanged. The ordered list is based on the encrypted transactions because, before the time $T$ has elapsed, the sequencing layer cannot solve the puzzle and decrypt the transactions.
    \item After time \( T \) has passed, the sequencer solves the puzzle and decrypts the transaction using the derived symmetric key.    
\end{enumerate}
Using ZK proof in Radius is to check the validity of the time-lock puzzle before trying to solve the puzzle; this allows to prevent waste of resources and DOS attacks which is called Practical Verifiable Delay Encryption (PVDE). In Radius, ZK proof is used to verify the validity of the time-lock puzzle before attempting to solve it; this allows to prevent computational resource waste and mitigate DOS attacks \cite{pvde, Radius2024}.



\subsubsection{Trusted Execution Environment}
\hfill \break

In this approach, validators must operate Trusted Execution Environments (TEEs). These TEEs can be integrated with the threshold encryption technique. The concept involves each network node possessing a TEE, such as Intel's Software Guard Extensions (SGX), and collectively maintaining an encryption key. Transactions are encrypted and sent to these TEEs, where they remain encrypted until they are included in a finalized block on the blockchain. The security of this method relies on the cost associated with compromising a specific hardware device \cite{ferveo, rondelet2023threshold}.
\noindent Node operators can run a TEE on their server if it supports SGX, or they can use a remote server with SGX support. SGX is an extension found in some Intel CPUs, providing a set of operations that enable the creation of a TEE, known as an enclave. An enclave is a protected region in memory and the CPU where data and code are isolated and accessible only within this secure area. The encryption key is generated and stored within the CPU. During execution, data and code are automatically decrypted by the CPU, processed, and then re-encrypted to maintain security \cite{sgx}.

\noindent Using the TEE method with a public-key cryptography scheme, users have access to the public key of the enclave. They encrypt their messages using this public key and send them to the enclave. The enclave functions like a private mempool, keeping pending transactions hidden from view. Inside the enclave, transactions are decrypted, executed, and then published on the blockchain.

\subsubsection{Discussion}
\hfill \break
Table \ref{tab:PPmethods} shows an overview of the three primary methods for mempool encryption. In this context, users are anyone who wants to send their transactions through these methods.
With Threshold encryption, trust is placed in a committee of key holders. Delay encryption, on the other hand, is trustless as each user possesses their own encryption key independent of any specific committee. However, when employing TEEs, trust is required in the hardware to safeguard secret keys.

\noindent Regarding security, in Threshold encryption, the system remains secure and functional as long as at least $\frac{2}{3}$ of key holders are honest. In Delay encryption, security hinges on the safeguarding of users' symmetric keys. When utilizing hardware, security is contingent upon factors such as access controls, secure memory, and tamper resistance of the hardware.

\noindent Each method faces distinct challenges. With Threshold encryption, the risk lies in committee members colluding to disclose private keys to front-runners or ceasing decryption of transactions. 
In delay encryption, users are responsible for generating puzzles for each transaction and safeguarding their symmetric keys. Consequently, users may face issues such as losing their keys or having them stolen. 
Challenges in hardware usage include compatibility with other software and modules, as well as reliance on specific hardware devices for our solution.

\noindent The complexity of the threshold encryption method can be high compared to other methods because it requires managing multiple keys and coordinating among participants, necessitating robust key distribution and management protocols. In delay encryption, the user must generate a time-lock puzzle and manage their symmetric key for each transaction. With TEE, all key management is handled by the hardware.
\begin{table*}
\caption{Prominent Mempool Encryption methods to prevent MEV-extraction}
\label{tab:PPmethods}
\begin{tabularx}{\textwidth} { 
  | >{\centering\arraybackslash}X 
  | >{\centering\arraybackslash}X  
  | >{\centering\arraybackslash}X 
  | >{\centering\arraybackslash}X 
  | >{\centering\arraybackslash}X | }
 \hline
 \textbf{Privacy-preserving methods} & \textbf{Trust Level} & \textbf{Security} & \textbf{Challenge} & \textbf{Key Management} \\ 
 \hline
 \hline
 Threshold Encryption  & Committee of key holders & 2/3 of key holders  & Collusion of Committee members & Committee members\\
\hline
 Delay Encryption  & Trustless  & Users' symmetric key & Generating puzzles \& keeping symmetric keys safe & User\\
\hline
 TEE & Hardware & Hardware  & Compatibility, Hardware dependency, Centralization issue & Hardware\\
\hline
\end{tabularx}
\end{table*}

\subsection{Private Transactions}
Private transactions are sent directly to validators, bypassing the public mempool. This approach enhances privacy by avoiding exposure to front-running attacks. The process of sending a private transaction is as follows \cite{empiricalprivate}.
\begin{enumerate}
    \item \textbf{Create the Transaction: }The user generates a transaction without any encryption, just like a standard Ethereum transaction.
    
    \item \textbf{Direct Submission to Validators: }Instead of broadcasting the transaction to the public mempool, the user sends the transaction directly to one or more specific validators.
    This can be done through Private Remote Procedure Call Endpoints (RPC) for Ethereum or using specialized services that facilitate private transactions (e.g., Flashbots API \cite{flashbots_private}).   
    \item \textbf{Validator Inclusion: }The targeted validator receives the private transaction and includes it in the next block they produce (with high priority). Since the transaction is not visible in the public mempool, it is less susceptible to front-running and other types of attacks that exploit transaction visibility.
    \item \textbf{Broadcasting: }The block containing the private transaction is broadcast to the entire network. At this stage, other validators can see the transactions, but it is too late to be front-run.
    \end{enumerate}
\section{Smart Contract-level Protection}\label{SCLP}
The goal of Smart Contract-level MEV extraction protection is to mitigate the potential for MEV extraction within the application layer while preserving the protocol of the underlying blockchain network. The challenge lies in designing smart contracts that can protect against MEV extraction without altering the fundamental protocol of the blockchain. Certain types of applications are particularly susceptible to front-running. These applications include: DEX, Gambling, Auction, Buying a Domain name, and rare NFTs.  

Solutions fall into two primary categories: the first operates entirely within smart contracts, utilizing methods such as commit-and-reveal \cite{Shayan} and batch processing, while the second category involves off-chain components.

\begin{figure}
\centerline{\includegraphics[width=\columnwidth]{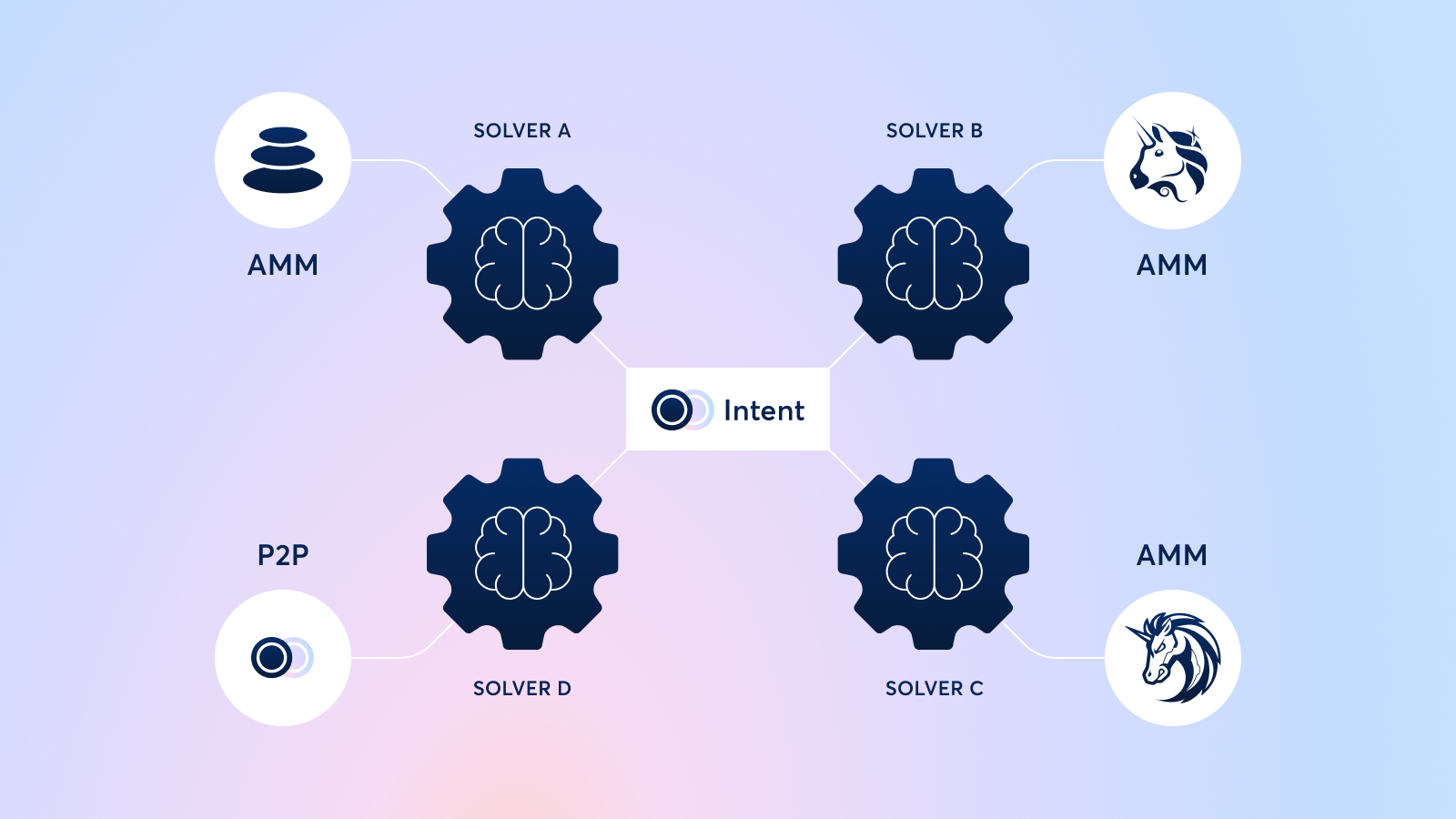}}
\caption{Cowswap Protocol Architecture. Adapted from \href{https://docs.cow.fi/cow-protocol/concepts/introduction/batch-auctions}{docs.cow.fi}}
\label{fig:cowswap}
\end{figure}

\textbf{CoWswap} \cite{CoWProtocol2024} is a solution in the second category; it is the first DEX aggregator, designed to facilitate asset swaps and limit orders with a hybrid MEV protection solution. 
CoWswap (see Fig. \ref{fig:cowswap}) sets up a private order flow marketplace where users can swap one asset for another at a specified price. Traders sign their trading intentions (intents), which are then aggregated off-chain and executed in batch settlements on-chain.
CoWswap introduces an uniform price clearing mechanism that assures traders receive the same price for an asset within the same block. It employs a decentralized network of solvers, or computer programs, that process the order-book to determine optimal prices and traded amounts. 
These solvers compete in batch auctions to find the best execution route across different liquidity sources, earning the right to execute the transactions and capturing any surplus from the trades \cite{CoWProtocol2024, Shoal2024}.

\begin{figure*}
\centerline{\includegraphics[width=\textwidth]{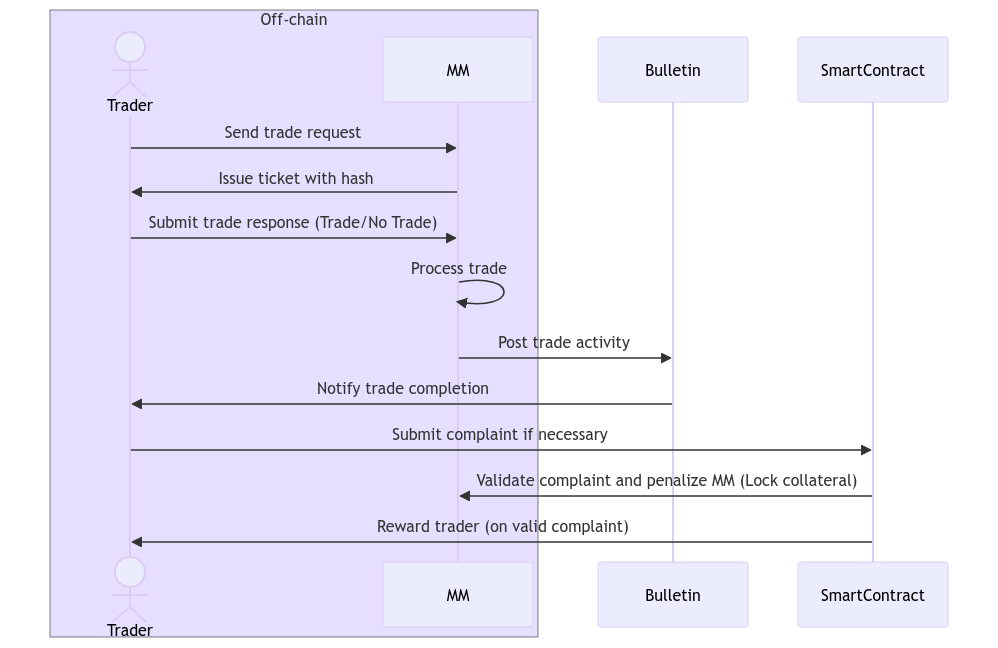}}
\caption{Fairmm Exchange Protocol Sequential Diagram.}
\label{fig:fairmm}
\end{figure*}

Ciampi et al. \cite{FairMM} propose an AMM cryptocurrency exchange named \textbf{FairMM}, which resists front-running attacks by incorporating off-chain components (second category). 
FairMM can prevent reordering trade transactions, censoring trades, and including specific trades request. The authors use game theory along with some incentive mechanisms in the smart-contract-level to remove reordering attacks. FairMM has higher throughput and lower trade cost compared to Uniswap which is an AMM crypto-market exchange. 
In FairMM, traders and the market maker (MM) communicate off-chain via secure channels like TLS. Traders form a queue, and the MM issues a ticket (identified by a cryptographic hash and signed by MM) to the trader at the front of the queue. This ticketing system ensures the integrity of the trading history and prevents reordering attacks. The process of FairMM is outlined in the following steps (see Fig. \ref{fig:fairmm}).
\begin{enumerate}
    \item A trader initiates a trade by sending a trade request, through an off-chain secure communication channel (e.g., TLS), to MM module which is responsible for managing and executing trades.
    \item MM issues a ticket to the trader. This ticket includes a cryptographic hash; it is signed by MM to ensure the integrity and authenticity of the request. MM manages a queue of trade requests. The ticketing system helps in maintaining the order of requests and prevents reordering attacks.
    \item The trader decides to proceed with or abort the trade and submit the response to MM
    \item MM processes trade requests by executing trades. This processing includes determining the appropriate price and executing the trade on behalf of the trader.
    \item MM posts the details of these trades to a public ledger which is called bulletin board. This action is on-chain and ensures transparency and accountability by making the trading activities publicly accessible.   
    \item The trader can submit any complaint if they find any missing or invalid ticket to a specific smart contract. The contract Handles complaints and enforces accountability by locking collateral and rewarding traders on valid complaints.       
    \end{enumerate}

\section{Proposer-Builder Separation}\label{PBS}
The primary objective of Proposer-Builder Separation (PBS) is to foster greater diversity among participants in blockchain networks by separating the roles of block proposers from those who actually create the blocks. This separation is intended to decentralize power and enhance network security by distributing responsibilities more widely among different parties. 

PBS can be implemented entirely on-chain, requiring changes to Ethereum's consensus layer, which have not yet been implemented. Currently, MEV-Boost \cite{MevBoost}, an off-chain implementation of PBS, is widely used, creating about $90\%$ of Ethereum's blocks under the Proof of Stake (PoS) consensus mechanism. 
This implementation specifically aims to mitigate issues related to MEV, promoting a more equitable and efficient transaction validation process by allowing separate entities to propose and construct blocks.
 The main idea is that instead of the block proposer trying to produce a revenue-maximizing block by themselves, they rely on an off-chain market where outside actors, called block-builders, produce bundles consisting of complete block contents and a fee for the proposer. 
 The proposer chooses the bundle with the highest fee. MEV-Boost allows validators to benefit from MEV without directly involving themselves in the MEV extraction process; this ensures a fairer distribution of rewards \cite{Flashbots2024, Flashbotsboost}. The process of MEVBoost is outlined in the following steps.
\begin{itemize}
    \item MEV searchers create a bundle of their profitable transactions from the Ethereum mempool. 
    \item MEV searchers send the bundles to a network of block builders in conjunction with a price bid to express their preferred position in a block.
     \item Block builders try to create the most profitable block with these bundles. The profitability of a block comes from both the MEV extracted from the transactions and the transaction fees. Builders are incentivized to include transactions that offer the highest MEV and fees
      \item After constructing the blocks, builders submit them to a relay, which checks their validity and calculates the total profit. This ensures that only valid and profitable blocks are added to the Ethereum blockchain. However, there is a risk of malicious behaviors from the relay, including potential block censoring or front-running.
    \item Block proposers or validators on the Ethereum network use MEV-Boost to request blocks from the relay; it selects and sends the most profitable block header (without payload) back to the validator. 
     \item Once the proposer signs the block header, the payload is released to the proposer. 
\end{itemize}

Since the block proposer signs the block header without knowing the payload or transaction content, they are unable to engage in malicious activities such as front-running or censoring transactions. 
Additionally, if a block proposer attempts to propose another block other than the one they signed previously, they risk being penalized by the network due to the possibility of double signing. 

Relays pose a centralization risk, as they have the potential for malicious actions such as censoring blocks or engaging in front-running activities. Currently there are a few relay providers that handles 90\% of the block production on Ethereum PoS \cite{RatedLabs2024}.
Providing incentives for running relay nodes can help prevent centralization. This could involve creating a token economy or other forms of compensation that encourage participation and discourage monopolization of relay services.
Establishing regular monitoring and auditing procedures for relay nodes can help detect and respond to attempts at centralization or abuse of power. This could involve both automated systems and manual reviews to ensure that relay nodes are operating in accordance with the principles of PBS \cite{relays2024}.

\begin{figure}
\centerline{\includegraphics[width=\columnwidth]{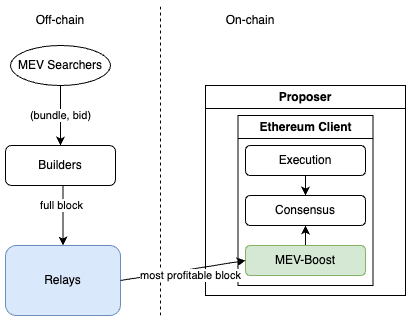}}
\caption{MEV-Boost Architecture}
\label{fig:mevboost}
\end{figure}

\section{Overview}\label{OVER}
Table \ref{tab:rollups} provides an overview of various approaches aimed at mitigating the MEV extraction issue. The majority of these solutions fall under the prevention category, focusing on either reducing or eliminating the opportunities for MEV exploitation. Among the solutions, PBS stands out as the only method designed to minimize the negative impacts by making MEV extraction more democratic. 

Implementing PBS through MEV-Boost does not add any latency to the network, as it operates off-chain primarily through relays. This ensures seamless operation without imposing additional load on the network. 
The overhead for the fair ordering method especially in a decentralized setting, all the nodes has to be agreed on an order of transactions which is predefined by the ordering policy of the network. So, it increases the need for communication between nodes. The algorithm load also increases with increasing number of sequencers engaging in the ordering process.
Privacy-preserving approaches generally involve encrypting the mempool, which can increase latency due to the encryption and decryption processes. 

The proposed mitigation strategies are generally applicable across various transaction types and are not limited to specific applications. However, the smart contract-level protection approach requires modifications to the application's design, making it impractical for already deployed smart contracts.

Both the privacy-preserving and PBS methods offer an additional advantage in terms of censorship resistance. Since validators do not have access to transaction details prior to their inclusion in a block, they lack any incentive to censor specific transactions. This enhances the fairness and integrity of transaction processing.
The other two strategies, smart contract-level protection and fair ordering, cannot prevent censorship by validators on their own. They need to incorporate additional methods, such as zero-knowledge proofs or slashing mechanisms, to effectively deter censorship.

Each method carries its own centralization risks. Fair ordering approaches might involve either a single sequencer or a permissioned group of sequencers. Privacy-preserving methods could encounter issues with a key management committee or dependencies on hardware devices. For PBS, the reliance on relay nodes introduces a potential point of centralization.

The fair ordering approach can be used on L2 chains without changing the Ethereum protocol. Privacy-preserving methods and PBS may need changes to the core protocol. This is particularly true if validators are involved in a key holder committee or if PBS is integrated directly into the network’s infrastructure.


While fair ordering policies can significantly mitigate MEV, they add complexity to the network protocol and require trust in the sequencers. These policies cannot entirely eliminate censorship, as sequencers might not prioritize transactions with high gas fees. Therefore, substantial incentives are needed to ensure that sequencers act honestly.

Rollup chains are moving towards decentralization, with some proposing to transition to a decentralized network of sequencers. However, this transition is costly for each Rollup chain, and with the emergence of new Rollup chains, incentivizing node operators to join each network becomes challenging. This leads to resource fragmentation.
A promising solution to these issues is the concept of shared sequencing. This approach envisions a decentralized network of sequencers that can be utilized collectively by various Rollup chains. Shared sequencing not only reduces the individual costs for each Rollup chain but also streamlines resource allocation, thereby preventing resource fragmentation and improving overall efficiency. By pooling resources and creating a unified network of sequencers, shared sequencing enhances the viability and scalability of decentralized Rollup chains.

Privacy-preserving methods, particularly those that enhance mempool privacy, show great promise in preventing front-running and ensuring censorship resistance. These methods can also be integrated with other strategies for mitigating MEV. Among these, delay encryption stands out as especially promising due to its trustless nature, which eliminates the need for a key management committee.
Future work on delay encryption involves making time-lock puzzles more secure and reducing the computational load on users. This will help improve the efficiency and usability of delay encryption, making it a more viable solution for enhancing privacy and security in blockchain networks.

There appears to be a strong incentive for validators, as only 10\% of blocks are built by validators not using MEV-Boost. Therefore, it is important to focus on addressing the challenges associated with the MEV-Boost solution, particularly by working towards the decentralization of relay networks and implementing incentives or slashing mechanisms to encourage honest behavior among relays.

\begin{table*}
\caption{A High-level Comparison Between Different MEV Mitigation Strategies }
\label{tab:over}
\begin{tabularx}{1\textwidth} { 
  | >{\centering\arraybackslash}X 
  | >{\centering\arraybackslash}X 
  | >{\centering\arraybackslash}X 
  | >{\centering\arraybackslash}X 
  | >{\centering\arraybackslash}X 
  | >{\centering\arraybackslash}X 
  | >{\centering\arraybackslash}X | }
 \hline
 \textbf{MEV Mitigation Strategies} & \textbf{Strategy} & \textbf{Latency} & \textbf{Application Dependent} & \textbf{Censorship Resistance} & \textbf{Centralization Risk} & \textbf{L1 Protocol Change}\\
 \hline
 \hline
 \textbf{Fair ordering}  & Prevention & Communication cost  & No & No & Sequencer nodes & No \\
\hline
 \textbf{Smart contract-level protection}  & Prevention  & Additional computation & Yes & No & --- & No\\
\hline
 \textbf{Privacy preserving}  & Prevention & Encryption/ Decryption & No  & Yes & Key management Committee/ Hardware & No/Yes \\
\hline
 \textbf{PBS}  & Democratizing  & --- & No & Yes & Relay nodes & No/Yes \\
\hline
\end{tabularx}
\end{table*}

 \section{Conclusion} \label{CON}
In conclusion, our research has yielded valuable insights into various MEV mitigation strategies applicable to Ethereum and L2 chains. As highlighted throughout the article, each approach presents its own set of challenges and advantages. Some methods offer security and efficiency but entail trust-related issues, while others aim to offload certain processes off-chain to reduce protocol overhead or mitigate MEV extraction side effects.

Notably, Mempool privacy utilizing Delay encryption and PBS emerges as promising approaches. Ongoing research in areas like \textbf{shared sequencing} \cite{sharedSequencing} and \textbf{Witness Encryption} \cite{stearn2022cryptographic} shows considerable potential. Further exploration into these topics is warranted, as new challenges may arise, necessitating continued investigation into the advancements made in these domains.

\raggedend
\printbibliography

@inproceedings{flashboys2.0,
  title={Flash boys 2.0: Frontrunning in decentralized exchanges, miner extractable value, and consensus instability},
  author={Daian, Philip and Goldfeder, Steven and Kell, Tyler and Li, Yunqi and Zhao, Xueyuan and Bentov, Iddo and Breidenbach, Lorenz and Juels, Ari},
  booktitle={2020 IEEE Symposium on Security and Privacy (SP)},
  pages={910--927},
  year={2020},
  organization={IEEE},
}

@inproceedings{Shayan,
  author = {Eskandari, Shayan and Moosavi, Seyedehmahsa and Clark, Jeremy}, 
  title = {SoK: Transparent Dishonesty: Front-Running Attacks on Blockchain},
  booktitle = {International Conference on Financial Cryptography and Data Security},
  year = {2019},
  volume = {11599},
  pages = {170-189},
}

@inproceedings{wendyGrowsUp,
  author = {Kursawe Klaus}, 
  title = {Wendy Grows Up: More Order Fairness},
  booktitle = {International Conference on Financial Cryptography and Data Security},
  year = {2021},
  pages = {191-196},
}

@inproceedings{FairMM,
  title={Fairmm: A fast and frontrunning-resistant crypto market-maker},
  author={Ciampi, Michele and Ishaq, Muhammad and Magdon-Ismail, Malik and Ostrovsky, Rafail and Zikas, Vassilis},
  booktitle={International Symposium on Cyber Security, Cryptology, and Machine Learning},
  pages={428--446},
  year={2022},
  organization={Springer},
}

@inproceedings{buyingtime,
  title={Buying Time: Latency Racing vs. Bidding for Transaction Ordering},
  author={Mamageishvili, Akaki and Kelkar, Mahimna and Schlegel, Jan Christoph and Felten, Edward W},
  booktitle={5th Conference on Advances in Financial Technologies (AFT 2023)},
  year={2023},
  organization={Schloss-Dagstuhl-Leibniz Zentrum f{\"u}r Informatik},
}

@inproceedings{wendy,
  title={Wendy, the good little fairness widget: Achieving order fairness for blockchains},
  author={Kursawe, Klaus},
  booktitle={Proceedings of the 2nd ACM Conference on Advances in Financial Technologies},
  pages={25--36},
  year={2020},
}

@article{ferveo,
  title={Ferveo: Threshold decryption for mempool privacy in bft networks},
  author={Bebel, Joseph and Ojha, Dev},
  journal={Cryptology ePrint Archive},
  year={2022},
}

@article{themis,
  title={Themis: Fast, strong order-fairness in byzantine consensus},
  author={Kelkar, Mahimna and Deb, Soubhik and Long, Sishan and Juels, Ari and Kannan, Sreeram},
  journal={Cryptology ePrint Archive},
  year={2021},
}

@article{sokPatric,
  title={Sok: Validating bridges as a scaling solution for blockchains},
  author={McCorry, Patrick and Buckland, Chris and Yee, Bennet and Song, Dawn},
  journal={Cryptology ePrint Archive},
  year={2021},
}

@article{xu2103sok,
  title={Sok: Decentralized exchanges (dex) with automated market maker (amm) protocols},
  author={Xu, Jiahua and Paruch, Krzysztof and Cousaert, Simon and Feng, Yebo},
  journal={ACM Computing Surveys},
  volume={55},
  number={11},
  pages={1--50},
  year={2023},
  publisher={ACM New York, NY},
}

@article{alam2024front,
  title={Front-running attack in decentralized finance in the metaverse: A systematic review},
  author={Alam, Tamimul and Ali, Md Asraf and Rahman, Md Hasibur and others},
  journal={International Journal of Science and Research Archive},
  volume={11},
  number={1},
  pages={2315--2324},
  year={2024},
  publisher={International Journal of Science and Research Archive},
}

@article{zhang2023combatting,
  title={Combatting Front-Running in Smart Contracts: Attack Mining, Benchmark Construction and Vulnerability Detector Evaluation},
  author={Zhang, Wuqi and Wei, Lili and Cheung, Shing-Chi and Liu, Yepang and Li, Shuqing and Liu, Lu and Lyu, Michael R},
  journal={IEEE Transactions on Software Engineering},
  year={2023},
  publisher={IEEE},
}

@article{balancer,
  title={A non-custodial portfolio manager, liquidity provider, and price sensor},
  author={Martinelli, Fernando and Mushegian, Nikolai},
  journal={URl: https://balancer. finance/whitepaper},
    year={2019},
}

@article{curve,
  title={Automatic market-making with dynamic peg},
  author={Egorov, Michael and Finance, Curve},
  journal={Retrieved Dec 2021},
  year={2021},
}

@article{uniswap,
  title={Uniswap v3 core},
  author={Adams, Hayden and Zinsmeister, Noah and Salem, Moody and Keefer, River and Robinson, Dan},
  journal={Tech. rep., Uniswap, Tech. Rep.},
  year={2021},
}

@article{rollupsHakim,
  title={Blockchain scaling using rollups: A comprehensive survey},
  author={Thibault, Louis Tremblay and Sarry, Tom and Hafid, Abdelhakim Senhaji},
  journal={IEEE Access},
  volume={10},
  pages={93039--93054},
  year={2022},
  publisher={IEEE},
}

@article{blindperm,
  title={Blindperm: Efficient mev mitigation with an encrypted mempool and permutation},
  author={Kavousi, Alireza and Le, Duc V and Jovanovic, Philipp and Danezis, George},
  journal={Cryptology ePrint Archive},
  year={2023},
}

@article{sharedSequencing,
  title={Shared Sequencing and Latency Competition as a Noisy Contest},
  author={Mamageishvili, Akaki and Schlegel, Jan Christoph},
  journal={arXiv preprint arXiv:2310.02390},
  year={2023},
}

@misc{dodoex,
  title = {What is DODO},
  author={DODO},
  url = {https://docs.dodoex.io/en/home/what-is-dodo},
  note = {Accessed: 2024-07-08},
}

@misc{stearn2022cryptographic,
  title={Cryptographic Approaches to Complete Mempool Privacy},
  author={Stearn, James},
  year={2022},
}

@inproceedings{verifiable,
  title={Verifiable random functions},
  author={Micali, Silvio and Rabin, Michael and Vadhan, Salil},
  booktitle={40th annual symposium on foundations of computer science (cat. No. 99CB37039)},
  pages={120--130},
  year={1999},
  organization={IEEE},
}

@inproceedings{kelkar2020order,
  title={Order-fairness for byzantine consensus},
  author={Kelkar, Mahimna and Zhang, Fan and Goldfeder, Steven and Juels, Ari},
  booktitle={Advances in Cryptology--CRYPTO 2020: 40th Annual International Cryptology Conference, CRYPTO 2020, Santa Barbara, CA, USA, August 17--21, 2020, Proceedings, Part III 40},
  pages={451--480},
  year={2020},
  organization={Springer},
}

@inproceedings{zhang2020byzantine,
  title={Byzantine ordered consensus without byzantine oligarchy},
  author={Zhang, Yunhao and Setty, Srinath and Chen, Qi and Zhou, Lidong and Alvisi, Lorenzo},
  booktitle={14th USENIX Symposium on Operating Systems Design and Implementation (OSDI 20)},
  pages={633--649},
  year={2020},
}

@inproceedings{heimbach2022sok,
  title={Sok: Preventing transaction reordering manipulations in decentralized finance},
  author={Heimbach, Lioba and Wattenhofer, Roger},
  booktitle={Proceedings of the 4th ACM Conference on Advances in Financial Technologies},
  pages={47--60},
  year={2022},
}

@inproceedings{delayencryption,
  title={Delay encryption},
  author={Burdges, Jeffrey and De Feo, Luca},
  booktitle={Annual International Conference on the Theory and Applications of Cryptographic Techniques},
  pages={302--326},
  year={2021},
  organization={Springer},
}

@inproceedings{DKGAlgorithm,
  title={Practical large-scale distributed key generation},
  author={Canny, John and Sorkin, Stephen},
  booktitle={International Conference on the Theory and Applications of Cryptographic Techniques},
  pages={138--152},
  year={2004},
  organization={Springer},
}

@inproceedings{Aequitas,
  title={Order-fairness for byzantine consensus},
  author={Kelkar, Mahimna and Zhang, Fan and Goldfeder, Steven and Juels, Ari},
  booktitle={Advances in Cryptology--CRYPTO 2020: 40th Annual International Cryptology Conference, CRYPTO 2020, Santa Barbara, CA, USA, August 17--21, 2020, Proceedings, Part III 40},
  pages={451--480},
  year={2020},
  organization={Springer},
}

@article{rondelet2023threshold,
  title={Threshold encrypted mempools: Limitations and considerations},
  author={Rondelet, Antoine and Kilbourn, Quintus},
  journal={arXiv preprint arXiv:2307.10878},
  year={2023},
}

@article{buildTimeLockEncryption,
  title={How to build time-lock encryption},
  author={Liu, Jia and Jager, Tibor and Kakvi, Saqib A and Warinschi, Bogdan},
  journal={Designs, Codes and Cryptography},
  volume={86},
  pages={2549--2586},
  year={2018},
  publisher={Springer},
}

@article{Helix,
  title={Helix: A fair blockchain consensus protocol resistant to ordering manipulation},
  author={Yakira, David and Asayag, Avi and Cohen, Gad and Grayevsky, Ido and Leshkowitz, Maya and Rottenstreich, Ori and Tamari, Ronen},
  journal={IEEE Transactions on Network and Service Management},
  volume={18},
  number={2},
  pages={1584--1597},
  year={2021},
  publisher={IEEE},
}

@article{surveyL2,
  title={A survey of Layer-two blockchain protocols},
  author={Gangwal, Ankit and Gangavalli, Haripriya Ravali and Thirupathi, Apoorva},
  journal={Journal of Network and Computer Applications},
  volume={209},
  pages={103539},
  year={2023},
  publisher={Elsevier},
}

@article{nitroArbitrum,
  title={Arbitrum Nitro: A Second-Generation Optimistic Rollup},
  author={Bousfield, Lee and Bousfield, Rachel and Buckland, Chris and Burgess, Ben and Colvin, Joshua and Felten, Edward W and Goldfeder, Steven and Goldman, Daniel and Huddleston, Braden and Kalodner, Harry and others},
  year={2018},
}

@article{F3B,
  title={F3b: A low-latency commit-and-reveal architecture to mitigate blockchain front-running},
  author={Zhang, Haoqian and Merino, Louis-Henri and Estrada-Galinanes, Vero and Ford, Bryan},
  journal={arXiv preprint arXiv:2205.08529},
  year={2022},
}

@article{choudhuri2024mempool,
  title={Mempool Privacy via Batched Threshold Encryption: Attacks and Defenses},
  author={Choudhuri, Arka Rai and Garg, Sanjam and Piet, Julien and Policharla, Guru-Vamsi},
  journal={Cryptology ePrint Archive},
  year={2024},
}

@article{empiricalprivate,
  title={An empirical study on ethereum private transactions and the security implications},
  author={Lyu, Xingyu and Zhang, Mengya and Zhang, Xiaokuan and Niu, Jianyu and Zhang, Yinqian and Lin, Zhiqiang},
  journal={arXiv preprint arXiv:2208.02858},
  year={2022},
}

@article{mastersthesisTobias,
  title={Scaling public blockchains—A comprehensive analysis of optimistic and zero-knowledge rollups},
  author={Schaffner, Tobias and Schaer, F},
  journal={A comprehensive analysis of optimistic and zero-knowledge roll-ups, University of Basel},
  year={2021},
}

@misc{relays2024,
  author = {EigenPhi},
  title = {From Censorship to Centralization: 4 Dimensions of Ethereum Relay's Public Goods Problem},
  year = {2024},
  url = {https://eigenphi.substack.com/p/eth-relay-public-goods-problem},
  note = {Accessed: 2024-07-20},
}

@misc{EspressoSys2024,
  author = {Espresso Systems},
  title = {Espresso Architecture: System Overview},
  year = {2024},
  url = {https://docs.espressosys.com/sequencer/espresso-architecture/system-overview},
  note = {Accessed: 2024-06-05},
}

@misc{espressoSequencer,
  author = {Espresso Systems},
  title = {Decentralizing Rollups: Announcing the Espresso Sequencer},
  year = {2022},
  url = {https://medium.com/@espressosys/decentralizing-rollups-announcing-the-espresso-sequencer-81c4c7ef6d97},
   note = {Accessed: 2024-07-02},
}

@misc{TimeBanditAttack,
  title = {Time Bandit Attack},
  author = {Mev Wiki},
  url = {https://www.mev.wiki/attack-examples/time-bandit-attack},
  note = {Accessed: 2024-07-02},
}

@misc{Metis2024,
 author = {Metis},
 title = {Decentralized Sequencer: Sequencer Node},
 year = {2024},
 url = {https://docs.metis.io/dev/decentralized-sequencer/overview/sequencer-node},
 note = {Accessed: 2024-04-26},
}

@misc{metisGOV2024,
 author = {Metis},
 title = {Governance Proposal: Decentralized Sequencer Governance},
 year = {2024},
 url = {https://ceg.vote/t/governance-proposal-decentralized-sequencer-governance/1922},
 note = {Accessed: 2024-04-26},
}

@misc{MevBoost,
  title = {MEV-Boost},
    author={Flashbots},
  url = {https://mevboost.pics/},
  note = {Accessed: 2024-07-19},
}

@misc{CoWProtocol2024,
 author = {CoW Protocol},
 title = {CoW Protocol Documentation},
 year = {2024},
 url = {https://docs.cow.fi/cow-protocol},
 note = {Accessed: 2024-04-23},
}

@misc{Shoal2024,
 author = {Shoal},
 title = {MEV Protection: DEX and Aggregator Anti-MEV Mechanisms},
 year = {2024},
 url = {https://www.shoal.gg/p/mev-protection-dex-and-aggregator},
 note = {Accessed: 2024-04-23},
}

@misc{flashbots_private,
  title = {Flashbots Auction Overview},
  author = {Flashbots},
  year = {2024},
  url = {https://docs.flashbots.net/flashbots-auction/overview},
  note = {Accessed: 2024-07-18},
}

@misc{Flashbots2024,
    author = {Flashbots},
    title = {Flashbots MEV Boost: Architecture Overview - Block Proposal},
    url = {https://docs.flashbots.net/flashbots-mev-boost/architecture-overview/block-proposal},
    year = {2024},
}

@misc{Flashbotsboost,
 author = {Mittal, Tarun},
 title = {Flashbots MEV Boost Explained in 5 Mins},
 year = {2024},
 url = {https://medium.com/roverx/flashbots-mev-boost-explained-in-5-mins-68e191140224},
 note = {Accessed: 2024-04-22},
}

@misc{RatedLabs2024,
 author = {Rated Labs},
 title = {Rated Network Explorer - Relays},
 year = {2024},
 url = {https://explorer.rated.network/relays?network=mainnet&timeWindow=all},
 note = {Accessed: 2024-04-22},
}

@misc{unlockingmempool,
 author = {Chaisomsri},
 title = {Unlocking Encrypted Mempools (1): Genesis and Blueprint},
 year = {2024},
 url = {https://medium.com/@chaisomsri96/unlocking-encrypted-mempools-1-genesis-blueprint-17f183508d72},
 note = {Accessed: 2024-04-22},
}

@misc{sgx,
  title = {Software Guard Extensions},
  author = {Wikipedia Contributors},
  year = {2024},
  url = {https://en.wikipedia.org/wiki/Software_Guard_Extensions},
  note = {Accessed: 2024-06-26},
}

@misc{encryptedmempools2023,
 author = {Charbonneau, Jon},
 title = {Encrypted Mempools},
 year = {2023},
 url = {https://joncharbonneau.substack.com/p/encrypted-mempools},
 note = {Accessed: 2024-04-21},
}

@misc{Radius2024,
 author = {Park, AJ},
 title = {Decentralizing Rollup Sequencers: Towards A Rollup-Centric Ethereum},
 year = {2024},
 url ={https://hackmd.io/@zeroknight/radius_decentralizing_rollup_sequencers#Order-Validation},
 note = {Accessed: 2024-04-21},
}

@misc{pvde,
  author = {ethresearch},
  title = {MEV-Resistant ZK-Rollups with Practical VDE/PVDE},
  year = {2024},
  url = {https://ethresear.ch/t/mev-resistant-zk-rollups-with-practical-vde-pvde/12677},
  note = {Accessed: 2024-06-25},
}

@misc{paradigm_optimistic,
  author = {Konstantopoulos, Georgios}, 
  title = {(Almost) Everything you need to know about Optimistic Rollup},
  type = {Paradigm Research},
  year = {2021},
  url = {https://www.paradigm.xyz/2021/01/almost-everything-you-need-to-know-about-optimistic-rollup},
}

@misc{shutter,
  title={Introducing Shutter -- Combating Front Running and Malicious MEV Using Threshold Cryptography},
  author={Shutter Network},
  year={2024},
  url={https://blog.shutter.network/introducing-shutter-network-combating-frontrunning-and-malicious-mev-using-threshold-cryptography/},
  note={Accessed: 2024-06-01},
}

\end{document}